\UseRawInputEncoding
\documentclass[pre,aps,twocolumn]{revtex4}

\usepackage{graphicx}
\usepackage{amsmath,empheq}
\usepackage{amssymb}
\usepackage{ifthen}
\usepackage{booktabs}

\usepackage{graphicx}
\usepackage{dcolumn}
\usepackage{bm}
\usepackage{longtable}
\usepackage{gensymb}

\newcommand{\bb}{\begin{equation}}
\newcommand{\ee}{\end{equation}}
\newcommand{\ba}{\begin{eqnarray*}}
\newcommand{\ea}{\end{eqnarray*}}
\newcommand{\rhor}{\rho({\bf r})}
\newcommand{\dd}{{\rm d}}
\newcommand{\rr}{{\mathbf r}}
\newcommand{\dr}{{\rm d}{\bf r}}

\begin{document}

\title{Height of a liquid drop on a wetting stripe}

\author{Alexandr \surname{Malijevsk\'y}}
\affiliation{ {Department of Physical Chemistry, University of Chemical Technology Prague, Praha 6, 166 28, Czech Republic;}
 {The Czech Academy of Sciences, Institute of Chemical Process Fundamentals,  Department of Molecular Modelling, 165 02 Prague, Czech Republic}}

\begin{abstract}
\noindent Adsorption of liquid on a planar wall decorated by a hydrophilic stripe of width $L$ is considered. Under the condition, that the wall is
only partially wet (or dry) while the stripe tends to be wet completely, a liquid drop is formed above the stripe. The maximum  height
$\ell_m(\delta\mu)$ of the drop depends on the stripe width $L$ and the chemical potential departure from saturation $\delta\mu$ where it adopts the
value $\ell_0=\ell_m(0)$. Assuming a long-range potential of van der Waals type exerted by the stripe, the interfacial Hamiltonian model is used to
show that $\ell_0$ is approached linearly with $\delta\mu$ with a slope which scales as $L^2$ over the region satisfying $L\lesssim \xi_\parallel$,
where $\xi_\parallel$ is the parallel correlation function pertinent to the stripe. This suggests that near the saturation there exists a universal
curve $\ell_m(\delta\mu)$ to which  the adsorption isotherms corresponding to different values of $L$ all collapse when appropriately rescaled.
Although the series expansion based on the interfacial Hamiltonian model can be formed by considering higher order terms, a more appropriate
approximation in the form of a rational function based on scaling arguments is proposed. The approximation is based on exact asymptotic results,
namely that $\ell_m\sim\delta\mu^{-1/3}$ for $L\to\infty$ and that $\ell_m$ obeys the correct $\delta\mu\to0$ behaviour in line with the results of
the interfacial Hamiltonian model. All the predictions are verified by the comparison with  a microscopic density functional theory (DFT) and, in
particular, the rational function approximation---even in its simplest form---is shown to be in a very reasonable agreement with DFT for a broad
range of both $\delta\mu$ and $L$.
 \end{abstract}

\maketitle

\begin{figure}
\includegraphics[width=8cm]{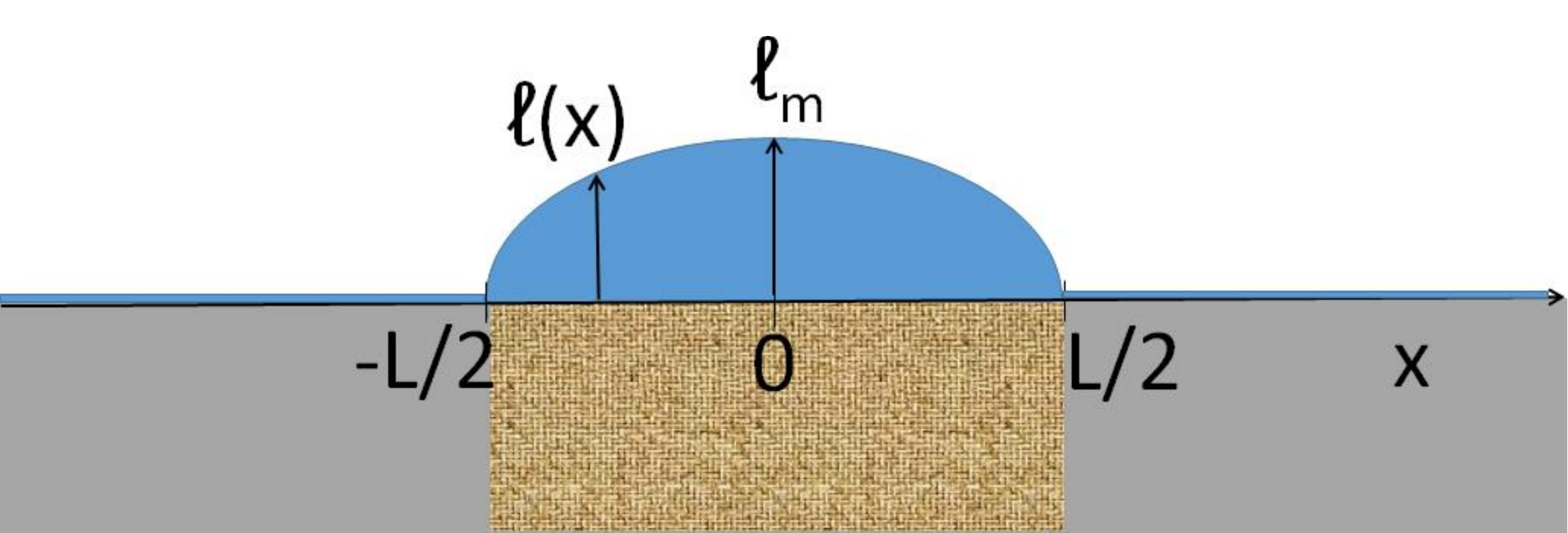}
\caption{A schematic two-dimensional projection of a liquid drop on a wetting stripe of width $L$. The remaining part of the wall is of a different
material and is only in a partially wetting state.}
\end{figure}

\section{Introduction}

It is well known that under certain conditions near the bulk liquid-gas coexistence, a macroscopic amount of liquid may intrude at the wall-gas
interface, making the wall completely wet  \cite{cahn, ebner, nakanishi, sullivan, dietrich, schick, bonn}. Similarly, in magnetic systems a
ferromagnet may be induced near a surface whose coupling with the spins is sufficiently strong, while the bulk remains disordered. For fluid systems,
the nature of adsorption depends on the thermodynamic path and also on the microscopic forces determining both fluid-fluid and wall-fluid
interactions between the atoms. In particular, at a fixed (subcritical) temperature $T>T_w$, where $T_w$ is the wetting temperature at which Young's
contact angle, made by the interface of the fluid coexisting phases and the wall, becomes zero, the mean height $\ell_\pi$ of the adsorbed liquid
film grows continuously and (in the absence of gravity) diverges as the bulk liquid-gas coexistence is approached from below. This critical
phenomenon is called complete wetting and is also characterized by a divergence of the parallel correlation length $\xi_{\parallel}$ along the
liquid-gas interface. The departure of the given thermodynamic state from the bulk liquid-gas coexistence can be quantified by the chemical potential
difference $\delta\mu=\mu_{\rm sat}(T)-\mu$ from its saturation value $\mu_{\rm sat}(T)$. In the presence of long-range microscopic forces, which
 asymptotically decay according to the power-law, the film thickness and the parallel correlation length diverge as \cite{lipowsky}
 \bb
 \ell_\pi(\delta\mu)\sim \delta\mu^{-\beta}\,,\label{wall}
 \ee
 and
  \bb
 \xi_{\parallel}(\delta\mu)\sim \delta\mu^{-\nu_{\parallel}}\,,\label{xi}
 \ee
as $\delta\mu\to0$, where the value of the critical exponents reflect the asymptotic behaviour of the microscopic forces. In particular, for
non-retarded dispersion forces, which can be modelled by the familiar Lennard-Jones potential, $\beta=1/3$ and $\nu_\parallel=2/3$ \cite{lipowsky}.

If the adsorbing wall is heterogenous, such that its  surface is modified chemically or geometrically, the wetting phenomena become considerably more
intricate in comparison with a homogenous and perfectly flat wall considered above. The process of complete wetting may then be accompanied by a
plenty of other interfacial phenomena such as filling \cite{hauge, rejmer, wood1, rejmer02, cone, bruschi, ancilotto, monson, parry_groove, our_prl,
our_wedge, our_groove, het_groove, rodriguez}, unbending \cite{santori, bauer2000, rascon2000}, depinning \cite{fang, mal19, mal20}, bridging
\cite{schoen, li, posp_bridge} and other morphological transitions \cite{lenz, bauer, gau, kargupta, krausch, wang, seemann, herminghaus, rauscher,
dokowicz, posp19}, whose interplay may give rise to very complex phase behaviour of the adsorbed fluid. In this work, let us consider a substrate
(wall) which is flat but decorated by a macroscopically long stripe of width $L$ which is of a material with a greater affinity to the liquid phase
than the rest of the wall, such that its wetting temperature $T_w^{\rm stripe}$ is lower than the wetting temperature of the wall $T_w^{\rm wall}$.
Therefore, considering a temperature $T$ which is in between the two wetting temperatures,  $T_w^{\rm stripe}<T<T_w^{\rm wall}$, the stripe will tend
to become completely wet, in contrast to the surrounding wall which is only in a partially wetting state. This means that sufficiently close to the
bulk phase coexistence, a liquid cylindrical drop forms above the stripe whose local height $\ell(x)$ has a maximum $\ell_m$ in the middle of the
stripe and is pinned to the wall at the stripe edges (see Fig.~1). The properties of liquid drops on heterogeneous walls have been studied previously
\cite{macdonald, jakubczyk4, jakubczyk6, jakubczyk7, trobo, posp17} at the bulk fluid coexistence in two \cite{jakubczyk6, jakubczyk7, trobo} and
three \cite{macdonald, jakubczyk4, posp17} dimensions. Systems involving only short-range forces have been shown to exhibit conformal invariance
which allows to predict the explicit form of the droplet height for a number of different domain shapes \cite{macdonald}; this is no more the case of
systems with the long-range forces which, however, still exhibit some scale invariance for the shape of the drop and its height which scales with the
width of the stripe as $\ell_m\propto\sqrt{L}$ \cite{posp17}.

The purpose of this work is to extend these studies by considering the growth of the drop as the saturation is approached from below. This means,
that we study the process of restricted, finite-size complete wetting and wish to know how the growth of the drop depends on two parameters:
$\delta\mu$ and $L$. To this end, we first briefly recapitulate the mean-field analysis based on the interfacial Hamiltonian model leading to the
simple formula for the height of the drop at saturation, $\delta\mu=0$. Next, the analysis is extended to the case of $\delta\mu>0$ which, however,
does not yield a solution in a closed form but can be expressed as a series in a dimensionless parameter proportional to $\delta\mu$. Although this
allows to obtain the asymptotic properties of the droplet growth upon approaching the saturation, the series expansion analysis is restricted only to
a vicinity of the saturation which thins with increasing the stripe width. Therefore, an alternative approach invoking finite-size scaling arguments
will be followed by taking advantage of the known droplet behaviour in the limits of $L\to\infty$ and $\delta\mu\to0$. This leads to a representation
of the droplet height in the form of a rational function of a single scaling parameter involving both $\delta\mu$ and $L$ with coefficients
determined by requiring a consistency with the series expansion for small $\delta\mu$ obtained previously from the interfacial Hamiltonian. In this
way, a very simple approximation for the droplet growth is found which, together with the series expansions up to the third order,  will be compared
with the results obtained from a classical density functional theory for a wide range of stripe widths.

\section{Analytical calculations}


On a mesoscopic level, a theoretical study of interfacial phenomena is based on an analysis of effective Hamiltonian models $H[\ell]$
\cite{lipowsky}, which describe the system state in terms of a shape $\ell(x)$ of the liquid-gas interface. This approach has been used with a great
success in the theory of wetting phenomena on a homogenous planar wall but has also been applied in various modified forms for other model
substrates. In this work, we will consider a simple interfacial model, which is defined by the following Hamiltonian per unit length
 \bb
H[\ell]=\int_{-L/2}^{L/2}\dd x\,\left[\frac{\gamma}{2}\left(\frac{\dd\ell(x)}{\dd x}\right)^2+W(\ell(x))\right]\,. \label{hl}
 \ee
In a close analogy with the analytical mechanics,  the effective Hamiltonian consists of two terms, where the first, ``kinetic'' contribution
proportional to the liquid-gas surface tension $\gamma$ is a free energy penalty for bending the interface. Here, we suppose that the interface
follows the symmetry of the wall and thus varies in only one dimension which is the Cartesian axis $x$ and is translation invariant along the stripe
corresponding to axis $y$. The second term of the functional is the so called binding potential $W(\ell)$ which describes the free-energy cost for
the presence of the liquid drop including the effect of microscopic forces.

At mean-field level, the equilibrium interface profile is found by simply minimizing (\ref{hl}), leading to the Euler-Lagrange equation
 \bb
\gamma\frac{\dd^2\ell}{\dd x^2}=W'(\ell)\,, \label{el}
 \ee
which is a subject of the boundary conditions $\ell(L/2)=\ell(-L/2)=0$ and $\frac{\dd\ell(0)}{\dd x}=0$. Here, an irrelevant but fixed microscopic
height of the interface at the stripe edges was set to zero. After integrating (\ref{el}) once and applying the boundary condition, one obtains
 \bb
\frac{\gamma}{2}\left(\frac{\dd\ell}{\dd x}\right)^2=W(\ell)-W(\ell_m)
 \ee
 where $\ell_m=\ell(0)$ is the maximum height of the drop. Further integration leads to the expression for the interface profile in the
 form of
 \bb
\int_{\ell(x)}^{\ell_m}\frac{\dd \ell'}{\sqrt{W(\ell')-W(\ell_m)}}=\sqrt{\frac{2}{\gamma}}|x|\,,\label{lx}
 \ee
which for $x=L/2$ yields an equation determining $\ell_m$:
 \bb
\int_{0}^{\ell_m}\frac{\dd \ell}{\sqrt{W(\ell)-W(\ell_m)}}=\frac{L}{\sqrt{2\gamma}}\,.  \label{lm}
 \ee

Eqs.~(\ref{lx}) and (\ref{lm}) are the general expressions for $\ell(x)$ and $\ell_m$, respectively, which follow exactly from the Euler-Lagrange
equation (\ref{el}). We now consider the specific but experimentally most relevant case, when the stripe interacts with the fluid via the long-range
dispersion forces. For a sufficiently large $L$ this implies the following form of the binding potential
 \bb
  W(\ell)=\delta\mu\Delta\rho\ell+\frac{A}{\ell^2}+\dots \label{bind}
 \ee
where the Hamaker constant $A>0$, since $T>T_w^{\rm stripe}$ and where the ellipses denote higher order terms in $1/\ell$. Here, the first term
linear in $\ell$ is the volume free-energy cost due to the occurrence of the metastable liquid, where $\Delta\rho=\rho_l(T)-\rho_v(T)$ is the density
difference between the liquid and gas particle densities of the bulk phases coexisting at the given temperature; the second term which decays as
$\ell^{-2}$ reflects the presence of the dispersion forces. When applied to a planar wall (corresponding to $L\to\infty$), in which case the l.h.s.
of (\ref{el}) is zero, the equilibrium film thickness $\ell_\pi$ is given by a minimization of the binding potential, which implies
 \bb
 \ell_\pi=\left(\frac{2A}{\delta\mu\Delta\rho}\right)^{\frac{1}{3}}\,, \label{ell_pi}
 \ee
as $\delta\mu\to0$, in line with the general result of Eq.~(\ref{wall}). One can also check that the value of the critical exponent for the parallel
correlation length according to the binding potential (\ref{bind}) has the expected value $\nu_{\parallel}=2/3$, since $\xi_{\parallel}\propto
\sqrt{1/W''}$ as follows from the Ornstein-Zernike theory \cite{schick}.

At saturation, i.e. for $\delta\mu=0$, the inverse quadratic form of the binding potential allows for a simple explicit solution for the droplet
height. In this case, a substitution of (\ref{bind}) into (\ref{lx}) leads to
 \bb
 \ell(x)=\sqrt{\ell_0^2-\frac{2A}{\gamma}\frac{x^2}{\ell_0^2}}
 \ee
  where the abbreviation for the the droplet maximum height at saturation $\ell_0\equiv\ell_m(\delta\mu=0)$ has been introduced, which itself satisfies
 \bb
\ell_0^2=\sqrt{\frac{A}{2\gamma}}L\,, \label{lm_dp0}
 \ee
 thus $\ell_0\propto \sqrt{L}$.

We now wish to extend these results below saturation, i.e. for $\delta\mu>0$. Substituting (\ref{bind}) to (\ref{lm}) leads after some rearranging to
 \bb
\frac{L}{\ell_m^2}\sqrt{\frac{A}{2\gamma}}=
  \int_0^{1}\frac{u\dd u}{\sqrt{1-u^2-2cu^2(1-u)}}
 \ee
 where $c\equiv \delta\mu\Delta\rho\ell_m^3/2A$, and using Eq.~(\ref{lm_dp0}) one obtains
 \bb
 \frac{\ell^2_0}{\ell^2_m}=
  \int_0^{1}\frac{u\,\dd u}{\sqrt{1-u^2-2cu^2(1-u)}} \label{integral}
 \ee
Note that with the help of Eq.~(\ref{ell_pi}) the dimensionless parameter $c$ can also be expressed  as $c=(\ell_m/\ell_\pi)^3$, hence  $0\le c<1$,
such that $c\to0$ as $\delta\mu\to0$ for $L$ fixed. Since we could not find the exact solution of  Eq.~(\ref{integral}) by evaluating the integral,
we proceed in a perturbative manner and expand the r.h.s. of Eq.~(\ref{integral}) around $c=0$ for which the solution is known and given by
Eq.~(\ref{lm_dp0}). Thus, we can write
 \bb
\frac{\ell_0^2}{\ell^2_m}=1+a_1c+a_2c^2+a_3c^3+\ldots\,, \label{series}
 \ee
 where the first coefficients are
  \begin{eqnarray}
 a_1&=&\frac{3}{4}\pi-2\,,\\
  a_2&=&12-\frac{15}{4}\pi\,,\\
   a_3&=&\frac{735}{32}\pi-72\,.
  \end{eqnarray}

 To lowest order (linear in $c$), Eq.~(\ref{series}) can be recast into
  \bb
  a_1\varepsilon x^5+x^2=1 \label{fifth}
  \ee
  with $x\equiv\ell_m/\ell_0$ and $\varepsilon\equiv\ell_0^3/\ell_\pi^3$.
  We seek the solution of Eq.~(\ref{fifth}) in the form of $x(\varepsilon)=1+x_1\varepsilon+\ldots$, which gives
  \bb
  x=1-\frac{a_1}{2}\varepsilon+\cdots\,,
  \ee
  hence we obtain that
   \bb
   \ell_m\approx\ell_0\left[1-\left(\frac{3}{8}\pi-1\right)\frac{\ell_0^3}{\ell_\pi^3}\right]\,. \label{linear}
   \ee

The linear (in $\varepsilon$ and thus in $\delta\mu$) form of Eq.~(\ref{linear}) assumes that the higher order terms of the expansion can be
neglected, which requires $\delta\mu$ small enough, such that $\ell_0<\ell_\pi$. Hence the condition, under which the linear approximation
(\ref{linear}) may be deemed reliable, can be expressed with the help of Eqs.~(\ref{ell_pi}) and (\ref{lm_dp0})  as
 \bb
 L<\frac{A^{1/6}\sqrt{2\gamma}}{(\Delta\rho\delta\mu)^\frac{2}{3}}\propto\xi_\parallel\,, \label{cond}
 \ee
 where the right hand side which diverges as $\delta\mu^{-2/3}$ for small $\delta\mu$ has been associated with the parallel correlation length.

The lowest order approximation given by Eq.~(\ref{linear}) is sufficient for a description of the asymptotic behaviour of the droplet height
approaching its saturation value  $\ell_0$. Thus, near the saturation, such that the condition (\ref{cond}) is satisfied, the dependence
$\ell_m(\delta\mu) $ is linear in $\delta\mu$ with a slope which scales quadratically with $L$. Furthermore, Eq.~(\ref{linear}) suggests a universal
behaviour of the droplet growth in the rescaled variables $\widetilde{\ell_m}=\ell_m/\sqrt{L}$ and $\widetilde{\delta\mu}=\delta\mu L^{3/2}$. We will
come back to this in section 4 where these predictions will be tested against a microscopic DFT.

One can expect that by accounting for higher order terms in expansion (\ref{series}) an improvement over the linear approximation (\ref{linear}) can
be achieved. For instance, up to the third order, Eq.~(\ref{fifth}) will be modified as follows
 \bb
 1=x^2+a_1x^5\varepsilon+a_2x^8\varepsilon^2+a_3x^{11}\varepsilon^3\,, \label{cubic}
 \ee
whose solution can be sought in the form of $x(\varepsilon)=1+x_1\varepsilon+x_2\varepsilon^2+x_3\varepsilon^3$, where the coefficients of the
expansion are $x_1=-a_1/2$ (reproducing Eq.~(\ref{linear})), $x_2=9/8a_1^2-a_2/2$, and $x_3=3a_1a_2-7/2a_1^3-a_3/2$. However, although the inclusion
of the higher order terms in the series expansion is expected to provide a better approximation compared to Eq.~(\ref{linear}), the condition
(\ref{cond}) must still be obeyed meaning that its performance worsens with $L$.


To obtain a more general approximation, one should seek an alternative representation of $\ell(\delta\mu)$. In fact, since a planar wall is the
natural reference system for our model, the approximate solution should be increasingly more reliable as the stripe width extends,  becoming exact in
the limit of $L\to\infty$, for which the planar solution $\ell_\pi(\delta\mu)$ should be reproduced. Note that this contrasts with the series
expansion representation, whose range of applicability shrinks with increasing $L$. Therefore, let us assume the solution  in the form of
 \bb
 \ell_m=\ell_\pi \Phi\left(\frac{\ell_0}{\ell_\pi}\right)\,, \label{scale}
 \ee
where the scaling function $\Phi$, which is a finite-size correction to complete wetting, satisfies the conditions: $\Phi(t)=1$, as $t\to\infty$ and
$\Phi(t)=t$ as $t\to0$. Alternatively, the scaling function could be written as $\tilde{\Phi}(L/\xi_\parallel)$, satisfying $\tilde{\Phi}(t)\propto
t^{1/2}$ as $t\to0$ but Eq.~(\ref{scale}) is more appropriate for the further analysis and suggests the following form of $\ell_m$:
 \bb
  \ell_m=\ell_\pi\frac{\sum_{i=1}^n a_it^i}{1+\sum_{i=1}^n b_it^i}\,, \label{rfa}
 \ee
 where $t=\ell_0/\ell_\pi$ and where the behaviour of the scaling function implies that $a_1=1$ and $a_n=b_n$.

Thus, requirements put on the correct behaviour of $\ell(\delta\mu)$ in the limit of $L\to\infty$ and for $\delta\mu=0$ imply that its form should
follow the rational function expansion (\ref{rfa}), which in the lowest order reduces to an additive rule for the reciprocals of all the present film
thicknesses: $\ell_m^{-1}=\ell_\pi^{-1}+\ell_0^{-1}$. However, such a remarkably simple expression does not comply with the correct linear behaviour
of $\ell_m(\delta\mu)$ near the saturation, which is the exact result of the interfacial Hamiltonian model in the limit of $\delta\mu\to0$.
Therefore, the correct asymptotic behavior (\ref{linear}) of the liquid growth is used as an additional constraint imposed on the rational function
approximation, which implies that its simplest form is of the third order:
  \bb
  \ell_m=\ell_\pi\frac{t+at^2+bt^3}{1+ct+dt^2+bt^3}\,. \label{rfa2}
 \ee
 The condition (\ref{linear}) then determines two of the  coefficients, namely that $b=d=a_1$. The other two coefficients $a$ and $c$ are undetermined,
 except that they must be equal. Since we do not impose any further conditions on the behaviour of $\ell_m(\delta\mu)$, they might be set to zero,
 which, moreover, guarantees the absence of the term proportional to $(\delta\mu)^{4/3}$ in the Taylor expansion of (\ref{rfa2}). Finally, we get
   \bb
  \frac{\ell_m}{\ell_0}=\frac{1+a_1t^2}{1+a_1t^2+a_1t^3}\,. \label{pade}
 \ee
 This is the rational function approximation for the droplet height which we propose and whose accuracy will be tested in section IV by a comparison
 with DFT.


\section{Density functional theory}

Classical density functional theory \cite{evans79} is a statistical mechanical tool to determine thermodynamic properties and correlation functions
of inhomogeneous fluids. It is based on the result that the grand potential of a molecular system is a functional of one-body particle density
$\rhor$ and its minimum with respect to $\rhor$ determines the equilibrium density profile and the thermodynamic free energy. The grand potential is
related to the intrinsic Helmholtz free energy ${\cal{F}}$  via the functional Legendre transform
  \bb
 \Omega[\rho]={\cal F}[\rho]+\int\dd\rr\rhor[V(\rr)-\mu]\,,\label{omega}
 \ee
where $\mu$ is the chemical potential and $V(\rr)$ is an external field experienced by each particle of the system.

Except for a few specific fluid models, the intrinsic free energy functional is not known exactly and must be thus approximated. The scheme of the
approximation depends on the model fluid; in particular, for simple fluids of the Lennard-Jones type as considered in this work, it is natural to
split the intrinsic free energy into several contributions in a perturbative manner \cite{hend}:
  \bb
{\cal F}[\rho]={\cal F}_{\rm id}[\rho]+{\cal F}_{\rm hs}[\rho]+{\cal F}_{\rm att}[\rho]\,. \label{fexp}
 \ee
 Here, ${\cal F}_{\rm id}$ is the ideal gas part due to purely entropic effects which is known exactly
   \bb
  \beta {\cal F}_{\rm id}[\rho]=\int\dr\rho(\rr)\left[\ln(\rhor\Lambda^3)-1\right]\,,
  \ee
where $\Lambda$ is the thermal de Broglie wavelength and $\beta=1/k_BT$ is the inverse temperature.

The remaining parts of the expansion (\ref{fexp}) describe the contribution to the free energy due to the fluid-fluid interaction. Here the fluid
pair interaction is described in the spirit of the Barker-Henderson perturbation 
theory \cite{barker}, such that it is split as follows
 \bb
 u(r)=u_{\rm HS}(r)+u_{\rm att}(r)\,,
 \ee
 where $u_{\rm HS}$ is the (reference) hard-sphere potential and $u_{\rm att}$ is the attractive tail:
  \bb
 u_{\rm att}(r)=\left\{\begin{array}{cc}
 0\,;&r<\sigma\,,\\
-4\varepsilon\left(\frac{\sigma}{r}\right)^6\,;& \sigma<r<r_c\,,\\
0\,;&r>r_c\,.
\end{array}\right.\label{ua}
 \ee
which is truncated at $r_c=2.5\,\sigma$ and where the parameter $\sigma$ is identified with the hard-sphere diameter.

The second term of the free-energy expansion (\ref{fexp}), ${\cal F}_{\rm hs}$, corresponds to the repulsive, hard-sphere part of the interatomic
interaction and is approximated  using Rosenfeld's fundamental measure theory (FMT) \cite{ros} according to which
 \bb
{\cal F}_{\rm hs}[\rho]=k_BT\int\dd\rr\,\Phi(\{n_\alpha(\rr)\})\,.
 \ee
Within FMT, the free energy density $\Phi$ depends on the set of weighted densities $\{n_\alpha\}$ which, within the original Rosenfeld approach,
consist of four scalar and two vector functions, which are given by convolutions of the density profile and the corresponding weight function:
 \bb
 n_\alpha(\rr)=\int\dr'\rho(\rr')w_\alpha(\rr-\rr')\;\;\alpha=\{0,1,2,3,v1,v2\}\,,
 \ee
where $w_3(\rr)=\Theta(R-|\rr|)$, $w_2(\rr)=\delta(R-|\rr|)$, $w_1(\rr)=w_2(\rr)/4\pi R$, $w_0(\rr)=w_2(\rr)/4\pi R^2$,
$w_{v2}(\rr)=\rr/R\delta(R-|\rr|)$, and $w_{v1}(\rr)=w_{v2}(\rr)/4\pi R$. Here, $\Theta$ is the Heaviside function, $\delta$ is Dirac's delta
function and $R=\sigma/2$.

Finally, the attractive free-energy contribution is treated on a mean-field level:
 \bb
 {\cal{F}}_{\rm att}[\rho]=\frac{1}{2}\int d{\bf{r}}_1\rho(\rr_1)\int d{\bf{r}}_2\rho(\rr_2)u_{\rm att}(|\rr_1-\rr_2|)\,.
 \ee

The external potential $V=V(x,z)$ can be split into the part induced by the wetting stripe which is formed by Lennard-Jones atoms interacting with
the fluid atoms via the potential
  \bb
   \phi_{\rm stripe}(r) = 4\varepsilon_w
  \left[
    \left(\frac{\sigma}{r}\right)^{12} - \left(\frac{\sigma}{r}\right)^6 \label{phi1}
  \right]\,,
 \ee
 with the strength parameter $\varepsilon_w$.

The remaining part of the wall is assumed to be completely dry, such that the wall atoms interact with the fluid atoms via the repulsive bit of the
Lennard-Jones potential:
 \bb
   \phi_{\rm wall}(r) = 4\varepsilon_w
    \left(\frac{\sigma}{r}\right)^{12}\,.
 \ee

The entire external potential is obtained by integrating the wall-fluid pair potential over the whole domain of the wall  which is assumed to be
formed by wall atoms that are distributed uniformly with a density $\rho_w$. Hence, the wall potential can be written as
 \begin{eqnarray}
  V(x,z)&=&\frac{4}{45}\pi\varepsilon_w\rho_w\sigma^3\left(\frac{\sigma}{z}\right)^9\nonumber+ V_{L}(x,z)\,, \label{v1}
 \end{eqnarray}
where the first term is the repulsive part of the Lennard-Jones $9$-$3$ potential, while the second part is the attractive contribution due to the
stripe of width $L$, which can be expressed as
 \begin{eqnarray}
V_{L}(x,z)&=&-4\varepsilon_w\sigma^6\rho_w\int_{x-L}^x\dd x'\int_{-\infty}^\infty\dd y'\int_{z}^{\infty}\dd z'\nonumber\\
 &&\times\frac{1}{(x'^2+y'^2+z'^2)^3}\label{v2}\\
  &=&\alpha_w\left[\frac{1}{(x-L)^3}-\frac{1}{x^3}+\psi_6(x-L,z)-\psi_6(x,z)\right]\nonumber
\end{eqnarray}
where
 \bb
\alpha_w=-\frac{1}{3}\pi\varepsilon_w\sigma^6\rho_w
 \ee
 and
 \bb
\psi_6(x,z)=-{\frac {2\,{x}^{4}+{x}^{2}{z}^{2}+2\,{z}^{4}}{2{z}^{3}{x}^{3} \sqrt {{x}^{2}+{z}^{2}}}}\,.
 \ee

The minimization of Eq.~(\ref{omega}) leads to the Euler-Lagrange equation for the density profile $\rhor$:
 \bb
 k_BT\ln[\rhor\Lambda^3]+\frac{\delta({\cal{F}}_{\rm hs}+{\cal{F}}_{\rm att})}{\delta\rhor}+\mu-V(\rr)=0\,, \label{el_dft}
 \ee
 which is solved iteratively on a two-dimensional grid $(x,z)$ in Cartesian coordinates using the Gaussian quadrature \cite{mal13}.

\section{Results}

\begin{figure}
\includegraphics[width=8cm]{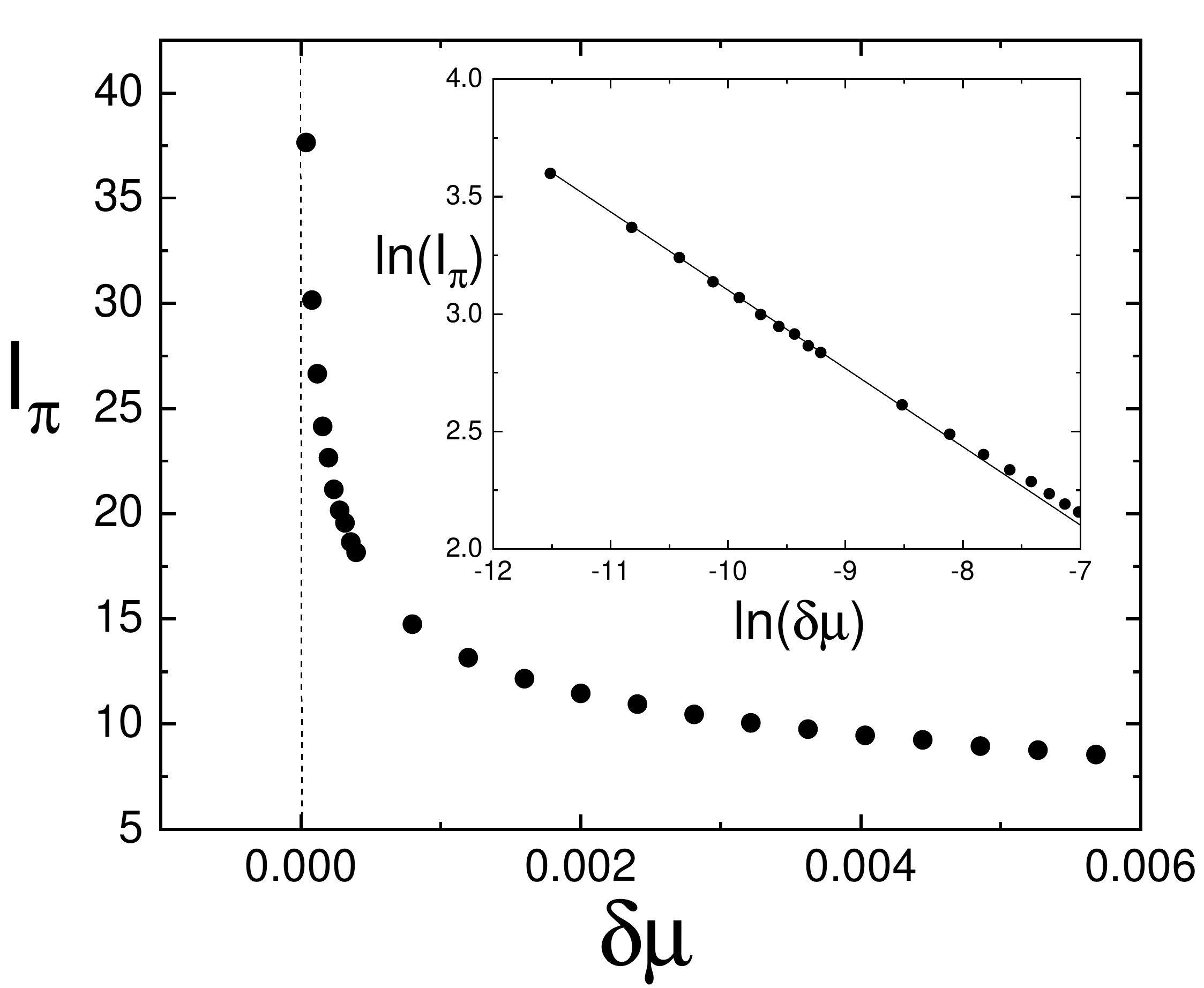}
\caption{A dependence of the thickness $\ell_\pi$ of the liquid film adsorbed on a planar wall on $\delta\mu=\mu_{\rm sat}-\mu$ at temperature
$T=0.92\,T_c$. The inset shows the behaviour of $\ell_\pi(\delta\mu)$ near the saturation as a log-log plot; here, the solid line is a linear fit to
the data with a slope of $-1/3$ which verifies the expected power-law dependence (\ref{ell_pi}).} \label{cw_plane}
\end{figure}

In this section the predictions obtained in section II are compared with the numerical DFT results. All the results correspond to
$\rho_w\varepsilon_w=1\cdot\varepsilon\sigma^{-3}$ and temperature $T=0.92\,T_c=1.15\,T_w$, where $T_c$ is the bulk critical temperature and $T_w$ is
the wetting temperature of the stripe. Since now on, all the length and energy quantities will be expressed in units of $\sigma$ and $\varepsilon$,
respectively.

Prior to discussing adsorption on a stripe of finite width $L$, let us first consider complete wetting on a homogeneous wall exerting Lennard-Jones
$9$-$3$ potential, which corresponds to the limiting case $L\to\infty$ of our substrate model. Fig.~\ref{cw_plane} displays a dependence of the film
thickness $\ell_\pi$ on the chemical potential departure from saturation, $\delta\mu$, as obtained from the DFT model formulated in  section III. For
a given value of $\delta\mu$, the equilibrium density profile $\rho(z)$ is obtained by solving Eq.~(\ref{el_dft}) with the external field
$V(z)=4\pi\varepsilon_w\rho_w\sigma^3\left[1/45\,(\sigma/z)^9-1/6\,(\sigma/z)^3\right]$, where $z$ is the distance from the wall. The film thickness
$\ell_\pi$ is determined from the density profile using the mid-density rule, $\rho(\ell_\pi)=(\rho_g+\rho_l)/2$. In order to accelerate the
iteration process near the saturation where $\ell_\pi$ grows rapidly, several liquid slabs of different widths were considered first and the one
corresponding to the minimal value of the approximate value of the grand potential obtained from (\ref{omega}) after just a few tens of iterations is
eventually used as the initial configuration for the full iteration process. In the inset of Fig.~\ref{cw_plane}, the graph is displayed as a log-log
plot, which shows a linear dependence with a slope of $-1/3$ for small $\delta\mu$, in accordance with the expected power-law dependence of
Eq.~(\ref{ell_pi}). Applying the linear regression to the log-log plot, the magnitude of the power-law has been estimated to be roughly $1.3$ which
allows to determine $\ell_\pi$ for further purposes; this value is consistent with that obtained directly from Eq.~(\ref{ell_pi}) by substituting for
$A=\pi/3\,\rho_w\varepsilon_w\sigma^6\Delta\rho$ \cite{schick}.

\begin{figure}
 \centerline{\includegraphics[width=8cm]{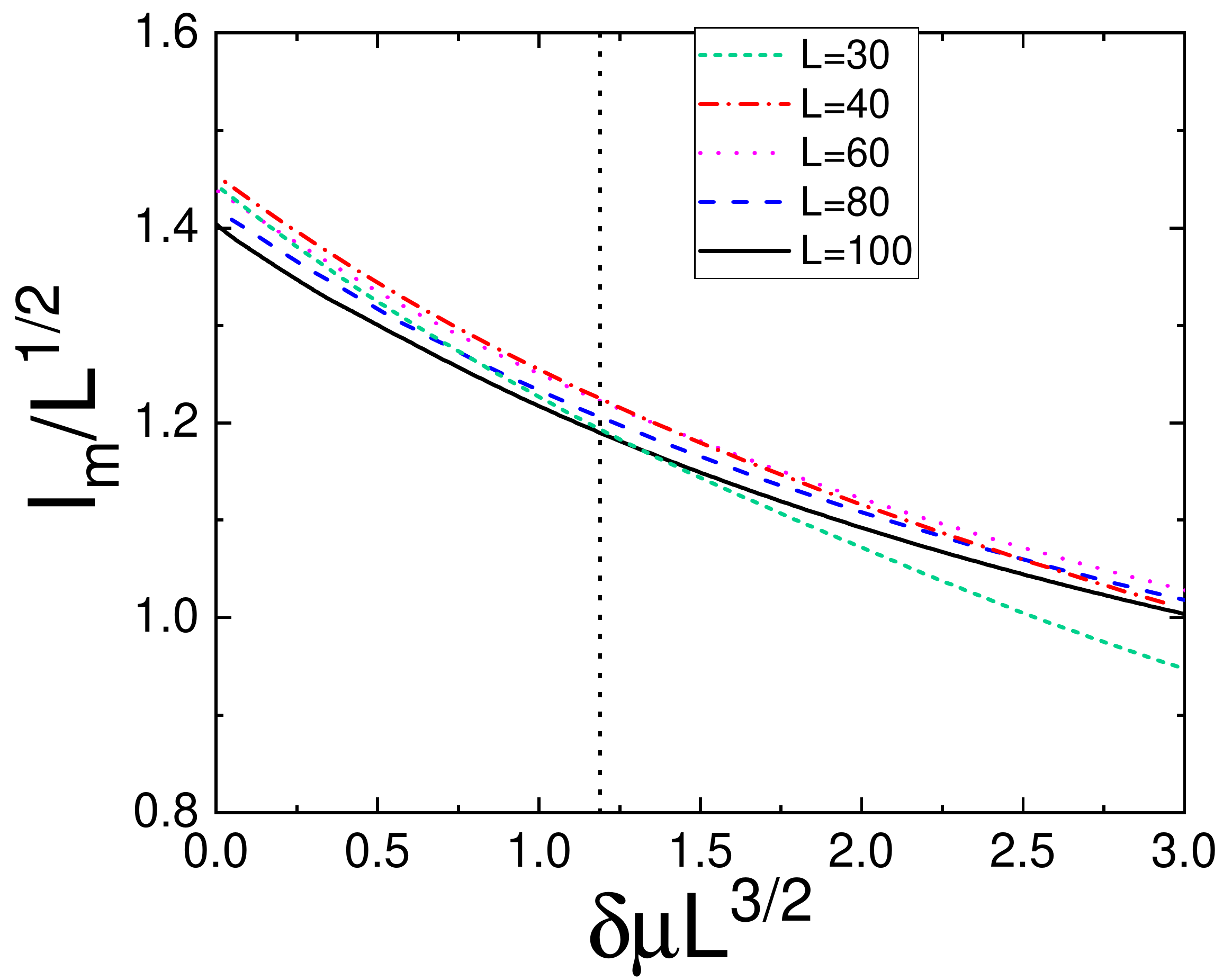}}
  \caption{DFT results showing a data collapse for the  dependence of $\ell_m$ on $\delta\mu$ for various values of
stripe widths $L$ in rescaled units. The dotted vertical line denotes an estimated borderline beyond which the scaling is not expected to be obeyed
anymore.} \label{rescaled}
\end{figure}

We now turn to complete wetting of a stripe of width $L$. First of all, we test the asymptotic results for the drop growth in the limit of
$\delta\mu\to0$ as given by Eq.~(\ref{linear}). According to this result, there exists a linear regime of $\ell(\delta\mu)$ sufficiently close to the
saturation which implies a universal behaviour of the drop growth in the rescaled variables $\widetilde{\delta\mu}=\delta\mu L^{3/2}$ and
$\widetilde{\ell_m}=\ell_m/\sqrt{L}$. In Fig.~\ref{rescaled}, the dependence of $\widetilde{\ell_m}(\widetilde{\delta\mu})$ is displayed for a number
of different stripe widths $L$ obtained from DFT. As can be seen, the data indeed almost collapse to a single curve confirming the scaling behaviour
for all the considered stripe widths up to the dotted line which corresponds to the value, for which $\ell_0\approx\ell_\pi$. Interestingly, the data
collapse extends even beyond this threshold except for the most microscopic case of $L=30$; this reflects the fact that the higher order coefficients
in the series expansion decay rather rapidly.

\begin{figure}
 \centerline{\includegraphics[width=8cm]{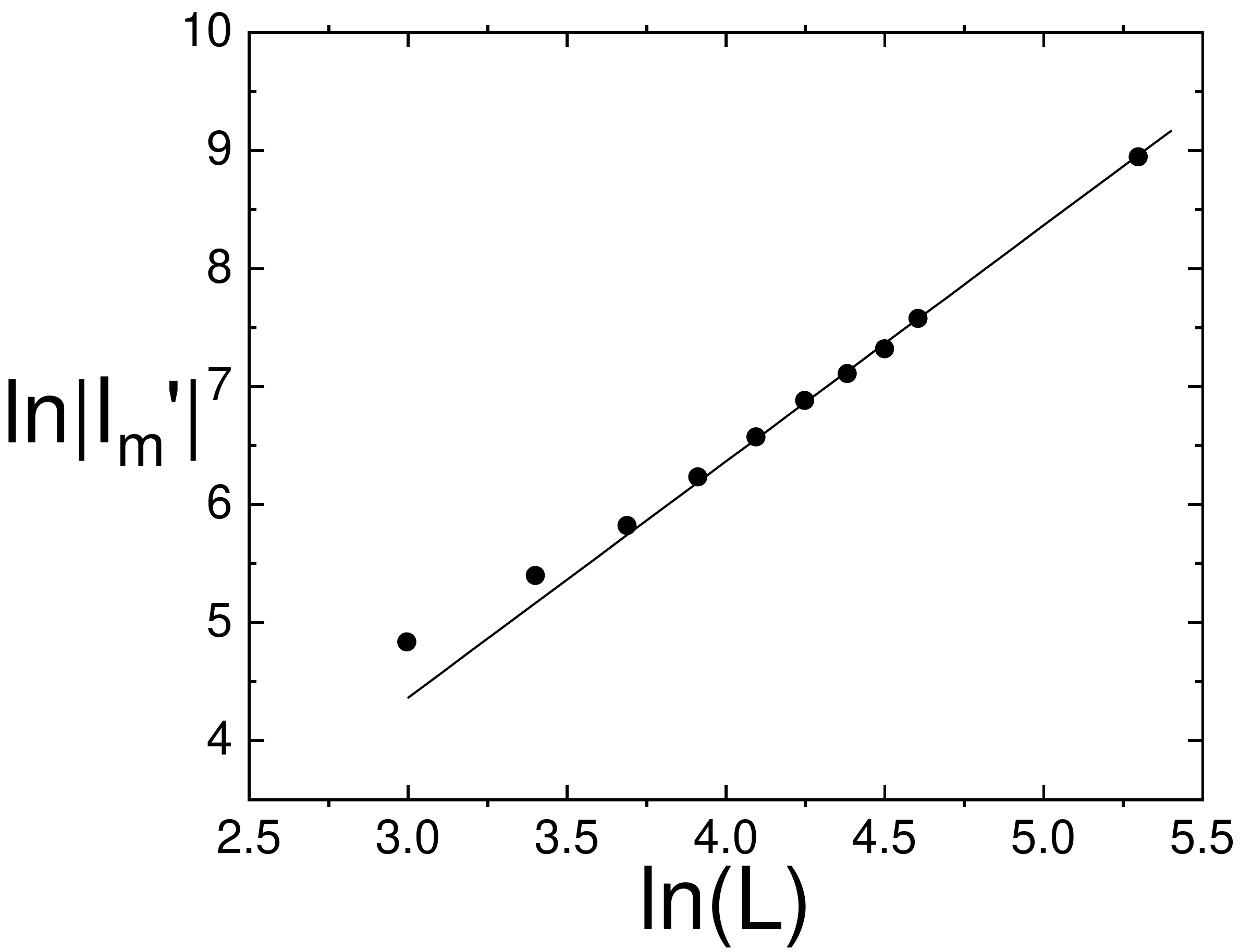}}
\caption{A log-log plot of the slope of $\ell_m(\delta\mu)$ at $\delta\mu=0$ for various stripe widths $L$ as obtained from DFT. For $L\ge40$ the
data follow the line with the slope of $2$ confirming that $\ell_m'(0)\propto L^2$ as follows from Eq.~(\ref{linear}). } \label{slope}
\end{figure}

A further test of the scaling properties of the drop growth near the saturation is shown in Fig.~\ref{slope}. Here, a modulus of the derivative of
$\ell_m$ with respect to $\delta\mu$ at $\delta\mu=0$ is displayed for various values of $L$. In the log-log plot, the dependence shows a near linear
behaviour and follows perfectly the line with a slope of $2$ for all the values of $L\ge40$, which provides a further support of the conclusions
drawn from Eq.~(\ref{linear}).

\begin{figure*}
\includegraphics[width=8cm]{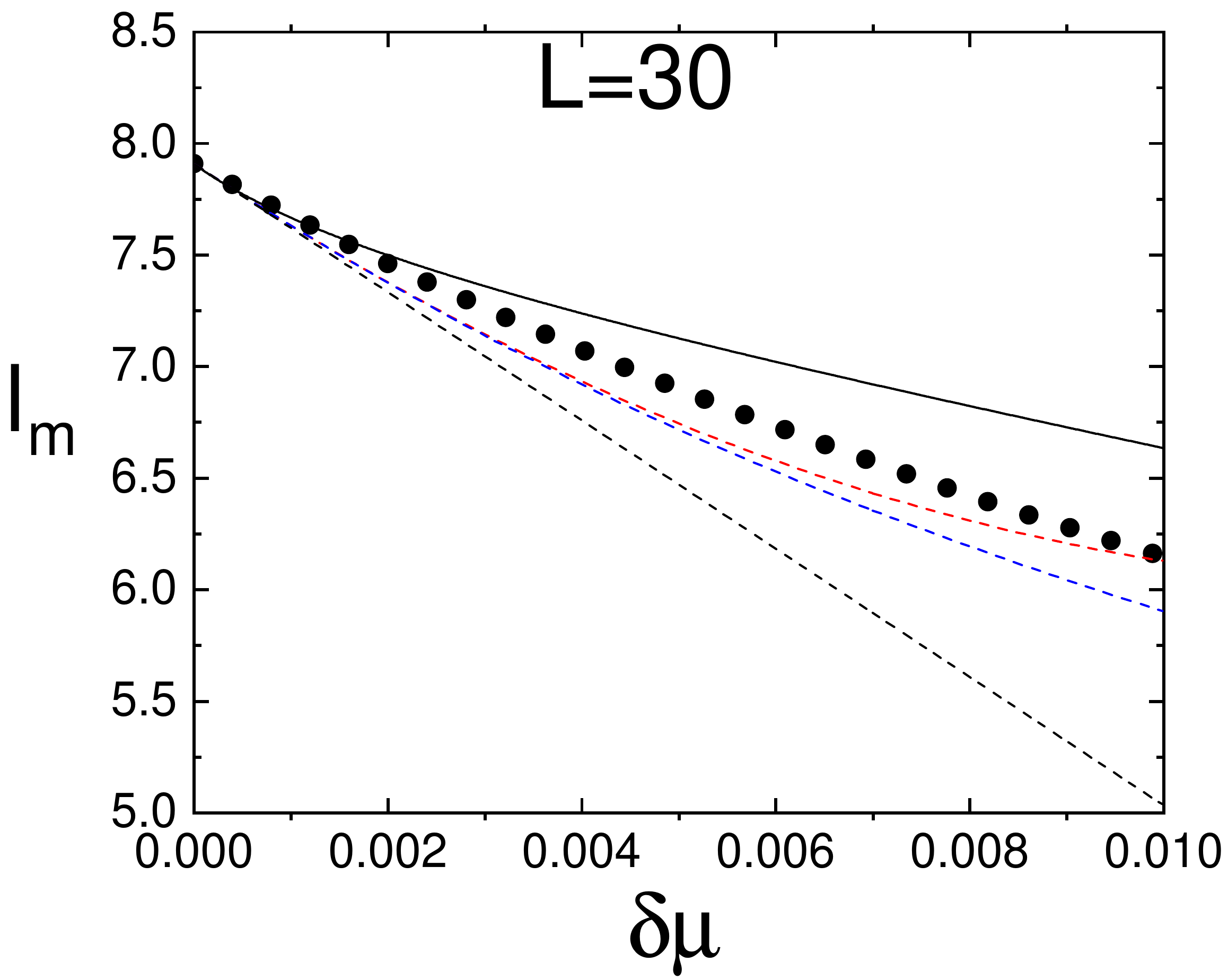} \includegraphics[width=8cm]{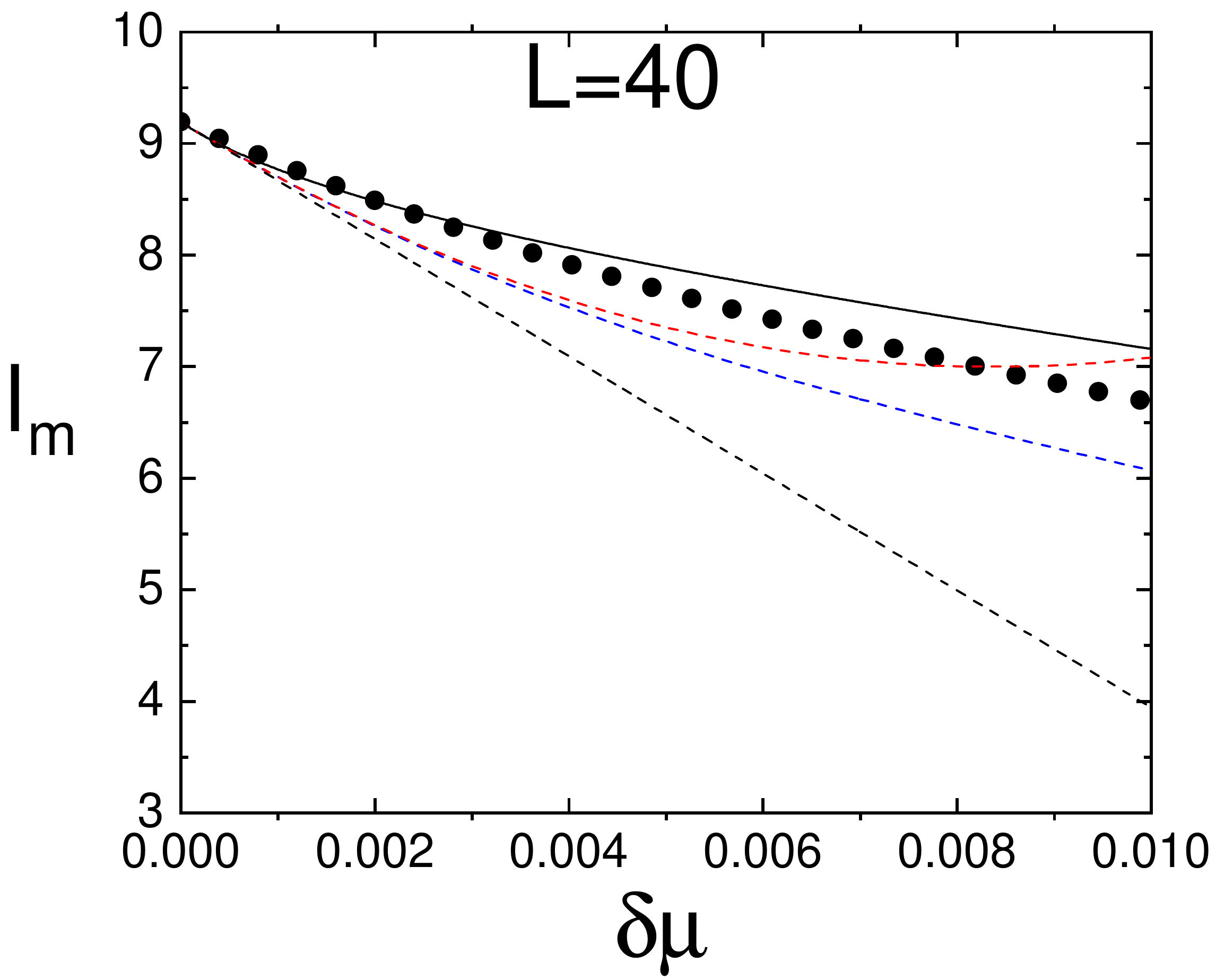}
\includegraphics[width=8cm]{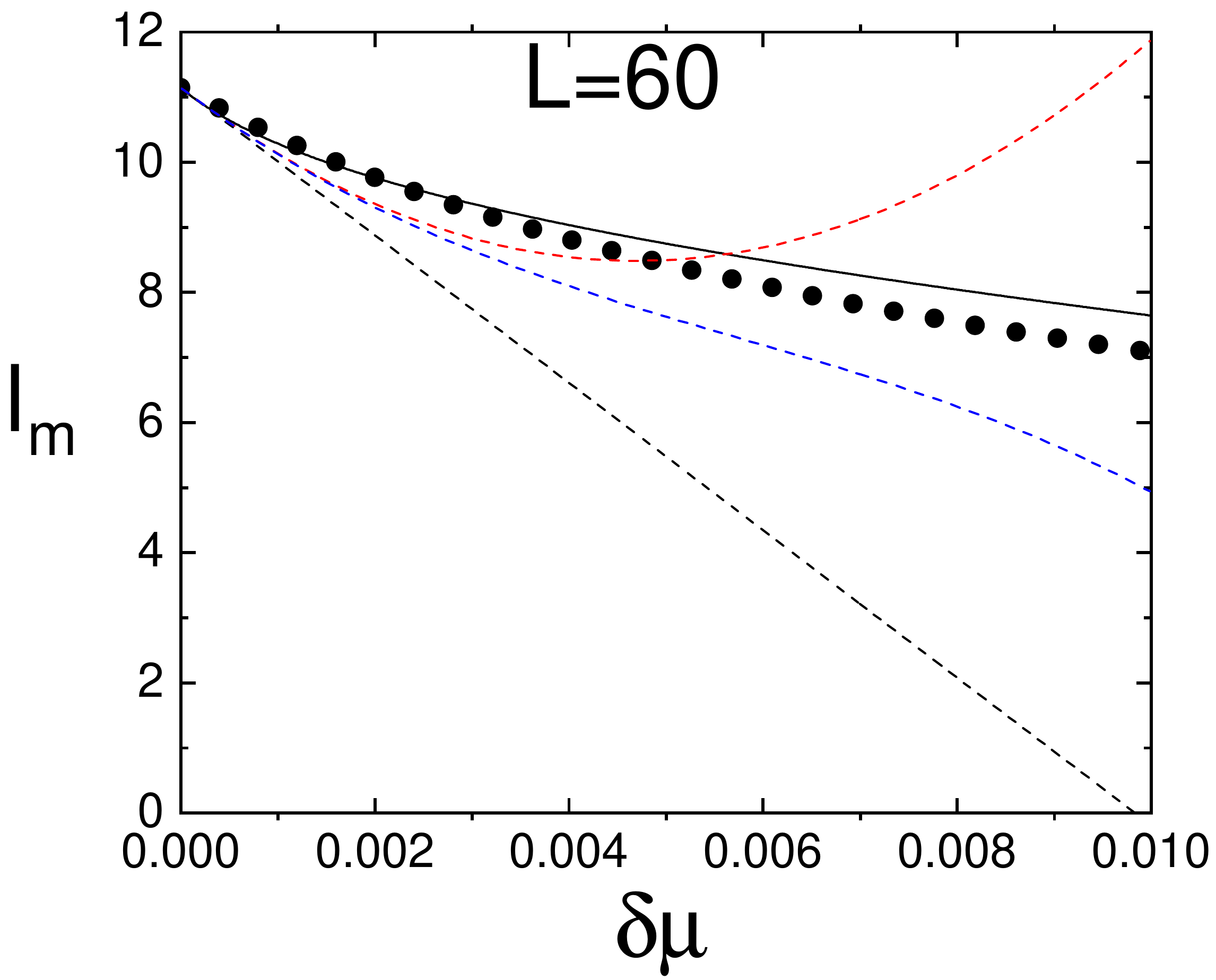} \includegraphics[width=8cm]{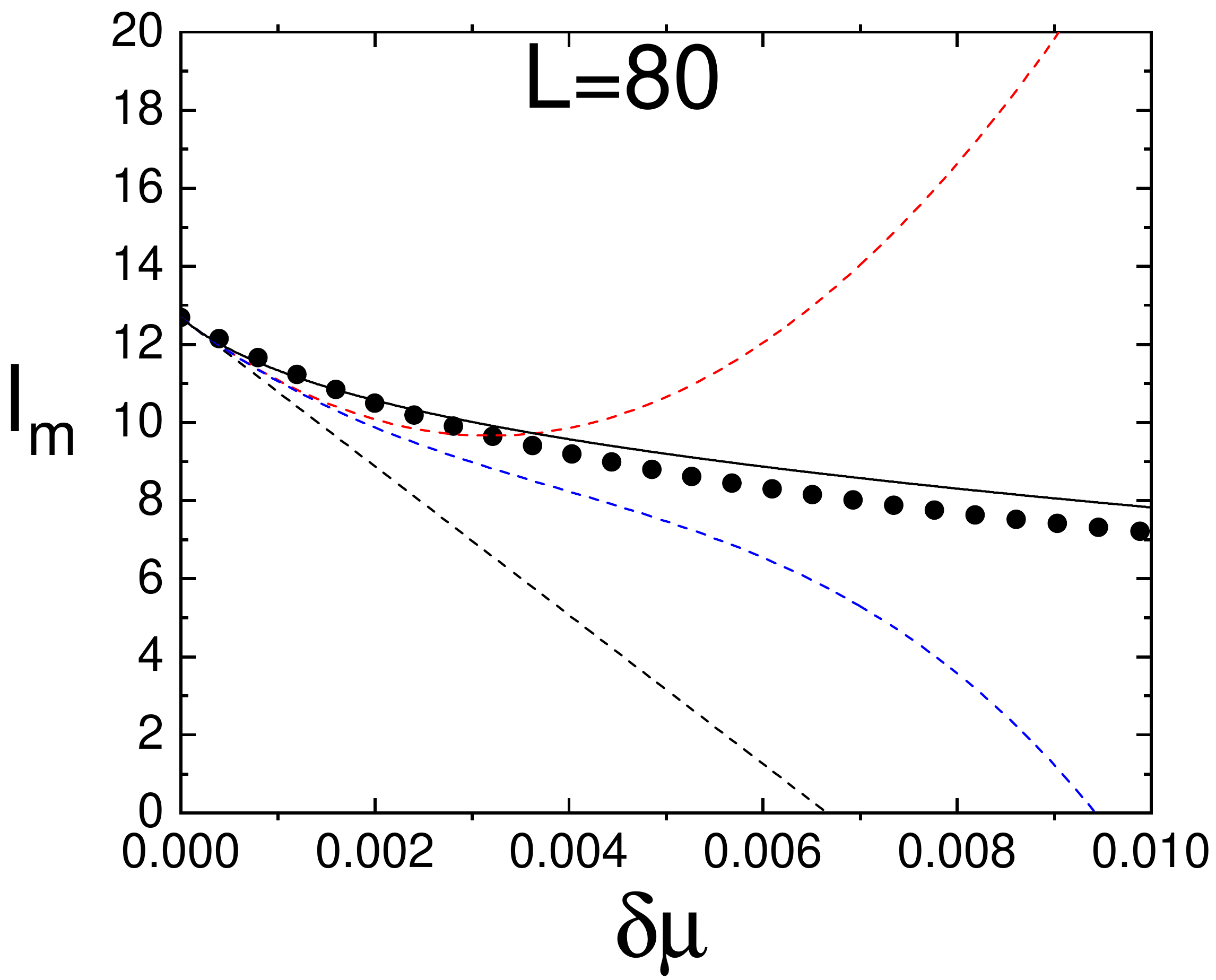}
\includegraphics[width=8cm]{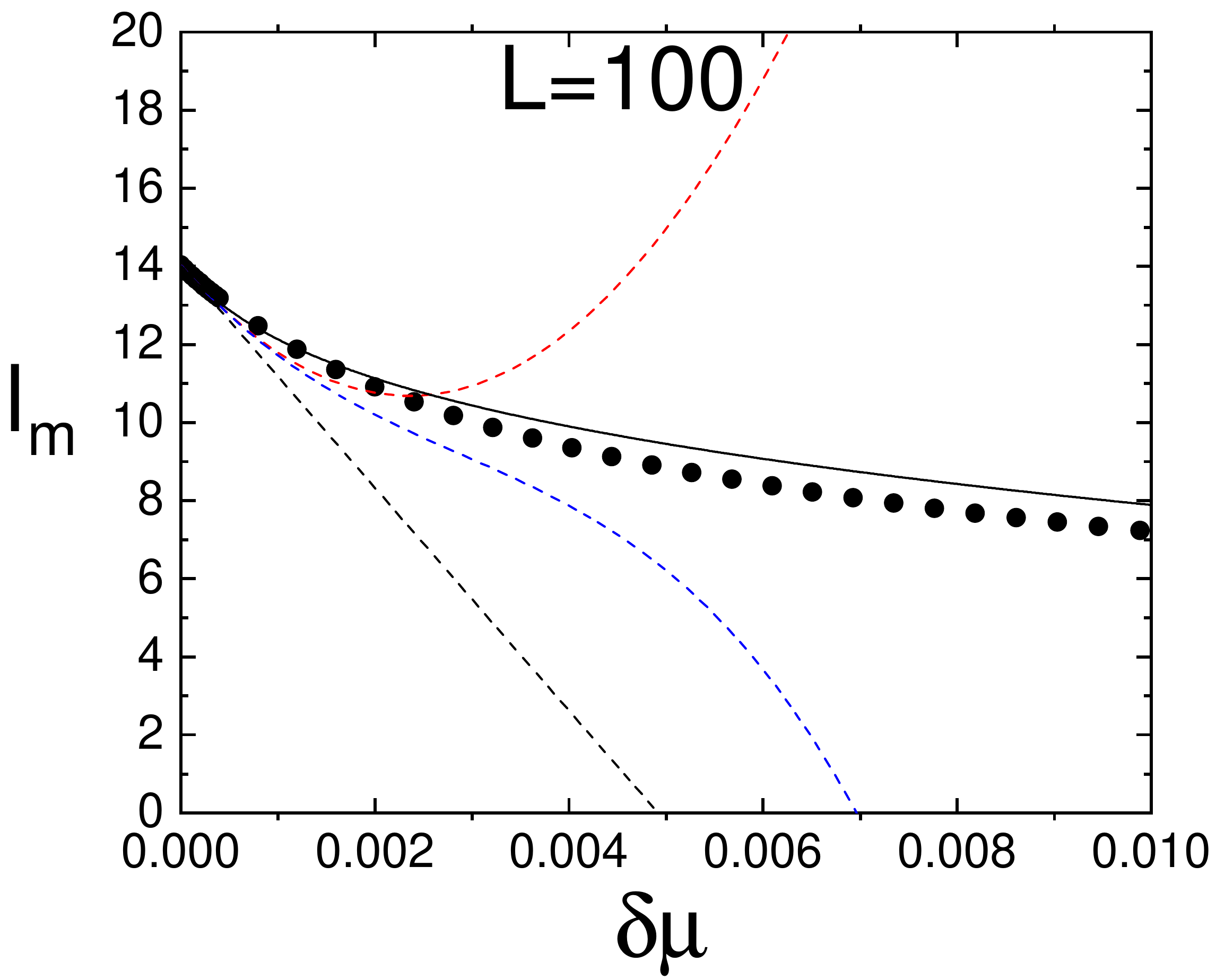} \includegraphics[width=8cm]{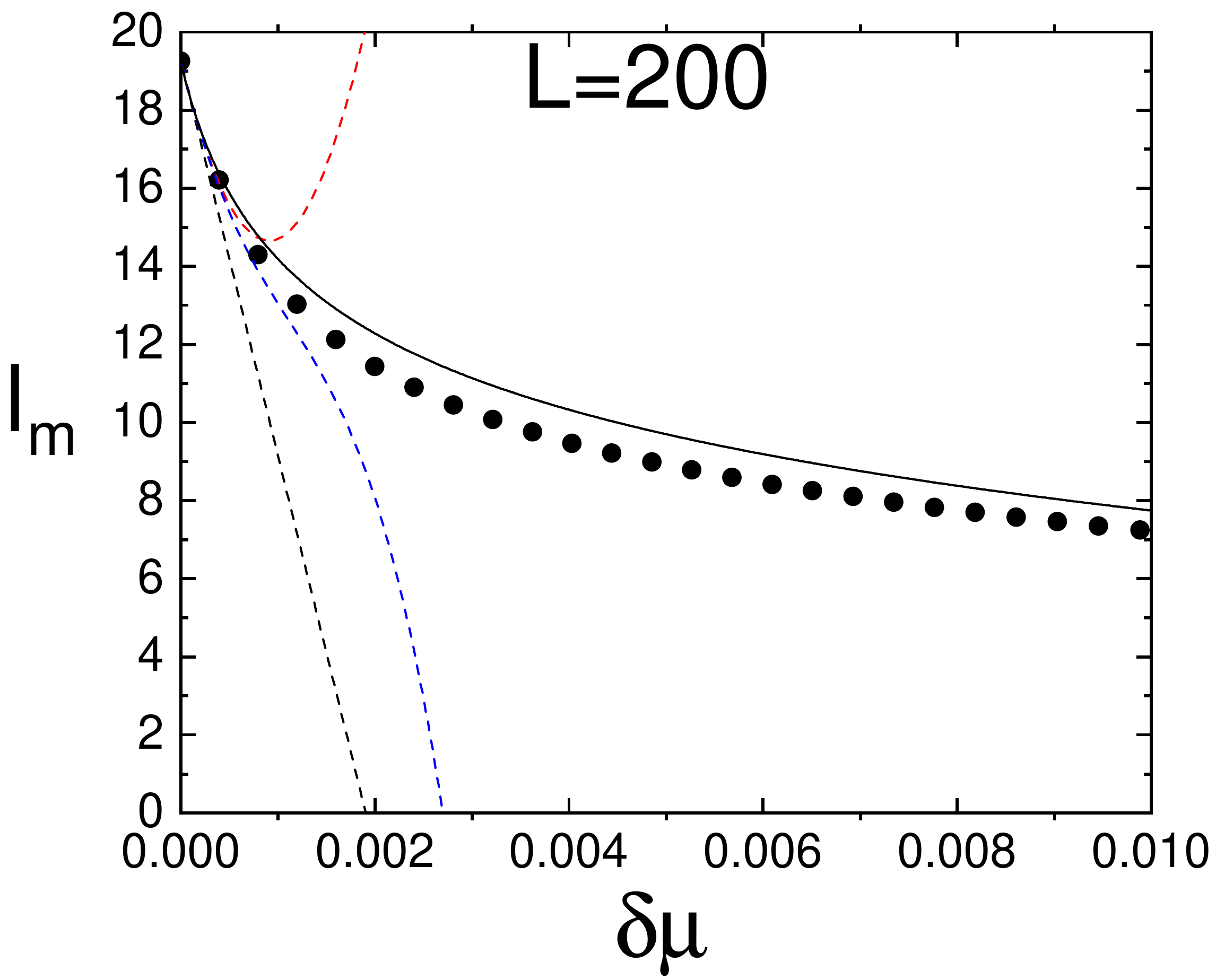}
\caption{A comparison between the DFT results (symbols) and the approximative theoretical predictions for various values of the stripe width $L$. The
dashed lines denote linear (black dashed), quadratic (red dashed) and cubic (blue dashed) orders of the expansion series (\ref{series}) as obtained
from the interfacial Hamiltonian model. The black solid line represents the rational function approximation (\ref{pade}).} \label{comparison}
\end{figure*}

Finally, let us compare the truncated series expansions and the rational function approximation with DFT over a large interval of $\delta\mu$, i.e.
even well below the saturation. In Fig.~\ref{comparison}, the comparison is made between results obtained from the series expansions up to the third
order as given by the solution of Eq.~(\ref{cubic}), the rational function approximation as given by Eq.~(\ref{pade}) and DFT for different values of
the stripe width. In all the cases, the saturation drop height $\ell_0$ as given by DFT has been used as the input to the theories, while the
power-law asymptotic form (\ref{ell_pi}) has been used to determine $\ell_\pi$. Since the scaling properties of Eq.~(\ref{linear}) has been verified
previously, it is not surprising already that the linear approximation captures properly the behaviour of $\ell_m(\delta\mu)$ near the saturation as
we can now see explicitly for all the cases. Moreover, we can also see that the linear regime thins with increasing of  $L$, which is in accordance
with condition (\ref{cond}). The quadratic and cubic approximations somewhat extend the interval of $\delta\mu$ over which the truncated series
expansion provides a reasonable agreement with DFT before they eventually blow up and in fact for the smallest $L$ the accuracy of the cubic
approximation is very reasonable over almost the entire range of the considered values of $\delta\mu$. However, the interval of applicability of the
higher order expansions also gets more and more narrow as $L$ increases, since the same competition between the stripe width and the parallel
correlation function (decreasing with $\delta\mu$) applies.
In contrast to the series expansions whose performance worsen by increasing both $\delta\mu$ and $L$, the simple rational function (\ref{pade})
provides a very reasonable approximation over the whole range of $\delta\mu$ for all the stripe widths; in general, the rational function
approximation slightly overestimates the DFT results but the discrepancy never exceeds one molecular diameter.

\section{Summary}

In this work, complete wetting of a macroscopically long stripe of width $L$ has been studied using mesoscopic and microscopic (DFT) methods. To this
end, a wall patterned by a macroscopically long stripe interacting with the fluid via a long-range potential has been considered and the process of
increasing the chemical potential (or pressure) towards the saturation was studied. Since the temperature of the system was fixed to the value
exceeding the wetting temperature of the stripe and the rest of the wall was considered to be non-wet (purely repulsive in DFT calculations), a
liquid drop is adsorbed at the stripe and its height $\ell_m$ grows continuously as the chemical potential is increased. The purpose of this study
was to describe the proces of the droplet growth $\ell(\delta\mu)$ as the saturation $\delta\mu=0$ is approached where the drop adopts its maximal
height $\ell_0\equiv\ell_m(0)$.

We first presented a mean-field analysis based on an interfacial Hamiltonian model; although  the corresponding Euler-Lagrange equation does not
provide an exact solution in a closed form for $\delta\mu>0$, it allows for the low $\delta\mu$ expansion around the saturation for which the exact
solution is known. From this it follows that $\ell_0$ is approached linearly in $\delta\mu$ with a slope scaling with $L^2$. These results suggest
that there exists a single universal curve of $\ell(\delta\mu)$ to which all adsorption isotherms collapse near the saturation when properly
rescaled, such that $\delta\mu\to\delta\mu L^{3/2}$ and $\ell_m\to \ell_m/\sqrt{L}$. This regime is restricted by a condition that the parallel
correlation function $\xi_\parallel$ pertinent to the stripe is larger than the stripe width $L$. All these predictions have been confirmed using a
microscopic DFT.

According to this analysis, the linear, as well as  any higher order regime obtained by truncating the series expansion shrinks with $L$, which is
verified by comparing the results with DFT for a broad range of stripe widths. The comparison has been made for the series expansions up to the third
order and although the higher order corrections improve the behaviour of $\ell_m(\delta\mu)$, such that the cubic approximation provides a fairly
good agreement with DFT results even well below saturation for stripe widths of few tens of molecular diameters, the series expansion clearly does
not represent the most appropriate form of $\ell_m(\delta\mu)$ for large $L$. Therefore, an alternative approach based on finite-size scaling
arguments has been applied requiring the correct behaviour of $\ell_m(\delta\mu)$ for macroscopically wide stripes, such that
$\ell_m(L,\delta\mu)\to\ell_\pi\sim\delta\mu^{-1/3}$ for $L\to\infty$. Together with a condition that $\ell_m(0)=\ell_0$ this suggests an expression
of $\ell_m$ in a form of a rational function of a single parameter $t=\ell_0/\ell_\pi$, which includes both $\delta\mu$ and $L$. These conditions
require that the degrees of the numerator and denominator polynomials forming the rational function are the same but apart from that there is no
other restriction. However, by imposing additionally that the low $\delta\mu$ behaviour of $\ell_m$ is consistent with the linear regime as
determined by the interface Hamiltonian model, it follows that the minimal degree of the rational function is three. Taking into account these
conditions, a Pad\'e approximant with a minimal number of terms that are necessary to reproduce the large $L$ and the small $\delta\mu$ behaviour of
$\ell_m$ is constructed. This approximation has been shown to be in a very reasonable agreement with the DFT results for the whole interval of the
considered stripe widths ranging from $30$ to $200$ molecular diameters and even far from the saturation.

Finally, let us note that the mean-field character of the analysis should not debase any of the conclusions made in view of the irrelevant effect of
the interfacial fluctuations in the considered three-dimensional substrate model \cite{lipowsky}, which thus should not disrupt the compactness of
the cylindrical droplets. Likewise, the pinning of the droplets at the edges of the stripe is expected to make them resistent towards the
Plateau--Rayleigh instability \cite{plateau, rayleigh}. Among many other potential extensions of this work, let us mention the possibility of the
prewetting jump in $\ell_m$ expected at lower temperatures and the anticipated finite-size shift in the prewetting line. The analysis could also be
extended over the saturation, i.e. for $\delta\mu$ negative, which would be relevant e.g. for slits formed of patterned walls with Young's contact
angle larger than $\pi/2$. The influence of the substrate geometry on the droplet growth on a wetting patch should also be elucidated. Most notably,
experimental verification of our results is desirable. Admittedly, this task is challenging and technically non-trivial. This is not only due to firm
requirements  of very accurate determination of the undersaturation but mainly because of the need to measure precisely the profile of drops of
molecular dimensions. This is in some contrast to macroscopically large drops (whose dimension is comparable with the capillary length) in which case
a crossover to the gravity dominated regime causing a drop flattening is already expected. However, the rapidly increasing development of the
nanotechnological methods promises a possibility of the experimental test of these predictions in a near future.

\begin{acknowledgments}
\noindent This work was financially supported by the Czech Science Foundation, Project No. GA 20-14547S.
\end{acknowledgments}


\begin{thebibliography}{99}

\bibitem{cahn}
 J. W. Cahn J. Chem. Phys.  {\bf 66}, 3667 (1977).

\bibitem{ebner}
C. Ebner and W. F. Saam, Phys. Rev. Lett. {\bf 88}, 1486 (1977).

 \bibitem{nakanishi}
H. Nakanishi and M. E. Fisher, Phys. Rev. Lett. {\bf 49}, 1565 (1982).

\bibitem{sullivan}
D. E. Sullivan and M. M. Telo da Gama, in {\it Fluid Interfacial Phenomena}, edited by C. A. Croxton (Wiley, New York, 1985).

\bibitem{dietrich}
S. Dietrich, in {\it Phase Transitions and Critical Phenomena}, edited by C. Domb and J. L. Lebowitz (Academic, New York, 1988), Vol. 12.

\bibitem{schick}
M. Schick, in {\it Liquids and Interfaces}, edited by J. Chorvolin, J. F. Joanny, and J. Zinn-Justin (Elsevier, New York, 1990).

\bibitem{bonn}
D. Bonn, J. Eggers, J. Indekeu, J. Meunier, and E. Rolley, Rev. Mod. Phys. {\bf 81}, 739 (2009).


\bibitem{lipowsky}
R. Lipowsky, Phys. Rev. Lett. {\bf 52}, 1429 (1984).


\bibitem{hauge}
E. H. Hauge, Phys. Rev. A {\bf 46}, 4994 (1992).

\bibitem{rejmer}
K. Rejmer, S. Dietrich, and M. Napirk\'owski, Phys. Rev. E {\bf 60}, 4027 (1999).

\bibitem{wood1}
A. O. Parry, C. Rasc\'on, and A. J. Wood, Phys. Rev. Lett. {\bf 83}, 5535 (1999).

\bibitem{rejmer02}
K. Rejmer, Phys. Rev. E {\bf 65}, 061606 (2002).

\bibitem{cone}
C. Rasc\'on and A. O. Parry, Phys. Rev. Lett. {\bf 94}, 096103 (2005).

\bibitem{bruschi}
L. Bruschi and G. Mistura, J. Low Temp. Phys.{\bf 157}, 206 (2009).

\bibitem{ancilotto}
F. Ancilotto, M. Barranco, E.S.  Hernandez, M. Pi, J. Low Temp. Phys.{\bf 157}, 174 (2009).

\bibitem{monson}
P. A. Monon, Microp. Mesopor. Mat. {\bf 160}, 47 (2012).

\bibitem{parry_groove}
C. Rasc\'on, A. O. Parry, R. N\"{u}rnberg, A. Pozzato, M. Tormen, L Bruschi, and G. Mistura, J. Phys.: Condens. Matter {\bf 25}, 192101 (2013).

\bibitem{our_prl}
A. Malijevsk\'y and A. O. Parry, Phys. Rev. Lett. {\bf 110}, 166101 (2013).

\bibitem{our_wedge}
A. Malijevsk\'y and A. O. Parry, J. Phys.: Condens. Matter {\bf 25}, 305005 (2013).

\bibitem{our_groove}
A. Malijevsk\'y and A. O. Parry, J. Phys.: Condens. Matter {\bf 26}, 355003 (2014).

\bibitem{het_groove}
A. O. Parry, A. Malijevsk\'y, and C. Rasc\'on, Phys. Rev. Lett. {\bf 113}, 146101 (2014).

\bibitem{rodriguez}
 A. Rodriguez-Rivas, J. Galv\'an, and J. M. Romero-Enrique, J. Phys. Condens. Matter {\bf 27}, 035101 (2015).

\bibitem{santori}
C. Rasc\'on, A. O. Parry, and A. Santori, Phys. Rev. E {\bf 59}, 5697 (1999).

\bibitem{bauer2000}
C. Bauer and S. Dietrich, Phys. Rev. E {\bf 61}, 1664 (2000).

\bibitem{rascon2000}
C. Rasc\'on and A. O. Parry, J. Phys.: Condens. Matter {\bf 12}, A369 (2001).

\bibitem{fang}
G. P. Fang and A. Amirfazli, Langmuir {\bf 28}, 9421 (2012).

\bibitem{mal19}
A. Malijevsk\'y, Phys. Rev. E {\bf 99}, 040801 (2019).

\bibitem{mal20}
A. Malijevsk\'y, Phys. Rev. E {\bf 102}, 012804 (2020).


\bibitem{schoen}
M. Schoen, Phys. Chem. Chem. Phys. {\bf 10}, 223 (2008).

\bibitem{li}
S. P. Li, Y. T. Chun, S. Zhao, H Ahn, D. Ahn, J. I.  Sohn, Y. B. Xu, P. Shrestha, M Pivnenko, D. P. Chu, Nat. Comun. {\bf 9}, 393 (2018).

\bibitem{posp_bridge}
A. Malijevsk\'y, A. O. Parry, and M. Posp\'\i \v sil, Phys. Rev. E {\bf 99},  042804 (2019).


\bibitem{lenz}
P. Lenz and R. Lipowsky, Phys. Rev. Lett. {\bf 80}, 1920 (1998).

\bibitem{bauer}
C. Bauer, S. Dietrich, and A. O. Parry,  Europhys Lett. {\bf 47}, 474 (1999).

\bibitem{gau}
H. Gau, S. Herminghaus, P. Lenz, and R. Lipowsky, Science {\bf 283}, 486 (1999).

\bibitem{kargupta}
K. Kargupta and A. Sharma, J. Chm. Phys. {\bf 116}, 3042 (2002).

\bibitem{krausch}
M. Geoghegan and G. Krausch, Prog. Polym. Sci. {bf 28}, 261 (2003).

\bibitem{wang}
J. Z. Wang, Z. H. Zheng, H. W. Li, W. T. S. Huck, and H. Sirringhaus, Nat. Mater. {\bf 3}, 171 (2004).

\bibitem{seemann}
R. Seemann, M. Brinkmann, E. J. Kramer, F. F. Lange, and R. Lipowsky, Proc. Natl. Acad. Sci. U.S.A. {\bf 102}, 1848 (2005).

\bibitem{herminghaus}
S. Herminghaus, M. Brinkmann, and R. Seemann, Annu. Rev. Mater. Res. {\bf 38}, 101 (2008).

\bibitem{rauscher}
S. Dietrich, M. Rauscher, and M. Napiorkowski, in {\it Nanoscale Liquid Interfaces}, edited by Ondar\c{c}uhu T. and Aim\'e J.-P. (Pan Stanford,
Singapore) 2013, p. 83.

\bibitem{dokowicz}
M. Dokowicz and W. Nowicki, Int. J. Heat Mass Transf. {\bf 115}, 131 (2017).

\bibitem{posp19}
M. Posp\'\i \v sil, M. L\'aska, and A. Malijevsk\'y, Phys. Rev. E {\bf 100}, 062802 (2019).

\bibitem{macdonald}
A. O. Parry, E.D. Macdonald, and C. Rasc\'on, J. Phys.: Condens. Matter {\bf 13}, 383 (2001).

\bibitem{jakubczyk4}
P. Jakubczyk and M. Napiorkowski, J Phys. Condens. Matter {\bf 16}, 6917 (2004).

\bibitem{jakubczyk6}
P. Jakubczyk, A. O. Parry, and M. Napiorkowski, Phys. Rev. E {\bf 74}, 031608 (2006).

\bibitem{jakubczyk7}
P. Jakubczyk and M. Napiorkowski, J Phys. A-Math. Theor. {\bf 40}, 2263 (2007).

\bibitem{trobo}
M. L. Trobo, E. V. Albano, and K. Binder, Phys. Rev. E {\bf 93},  052805 (2016).

\bibitem{posp17}
A. Malijevsk\'y, A. O. Parry, and M. Posp\'\i \v sil, , Phys. Rev. E {\bf 96}, 032801 (2017).



\bibitem{evans79}
R. Evans, Adv. Phys. {\bf 28}, 143 (1979).

\bibitem{hend}
R. Evans in {\it Fundamentals of Inhomogeneous Fluids}, (New York: Dekker  1992).

\bibitem{barker}
J. A. Barker and D. Henderson, J. Chem. Phys. {\bf 47}, 4714 (1967).

\bibitem{ros}
Y. Rosenfeld,  Phys. Rev. Lett. {\bf 63}, 980 (1989).

\bibitem{mal13}
A. Malijevsk\'y, J. Phys.: Cond. Matter {\bf 25}, 445006 (2013).

\bibitem{plateau}
J.~A.~F. Plateau, {\em Statique Expérimentale et Théorique des Liquides Soumis Aux
  Seules Forces Moléculaires, vol. 2}, Gauthier-Villars, (1873).

\bibitem{rayleigh}
L. Rayleigh, Proc. Lond. Math. Soc. {\bf s1-10}, 4 (1878).



\end{thebibliography}
\end{document}